\documentclass{article} 
\usepackage[preprint]{spconf}
\pdfoutput=1

\usepackage[utf8]{inputenc} 
\usepackage[T1]{fontenc}    
\usepackage{hyperref}       
\usepackage{url}            
\usepackage{booktabs}       
\usepackage{amsfonts}       
\usepackage{nicefrac}       
\usepackage{microtype}      
\usepackage{xcolor}         
\usepackage{listings}       
\usepackage{courier}
\usepackage{color}

\usepackage{graphicx}
\usepackage{makecell}
\usepackage{multirow}
\usepackage{multicol}
\usepackage{amsmath}

\title{TorchAudio 2.1: Advancing speech recognition, self-supervised learning, and audio processing components for PyTorch}

\name{
  \begin{tabular}{c}
  \it Jeff Hwang$^{*1}$, Moto Hira$^{*1}$, Caroline Chen$^{*\dagger}$, Xiaohui Zhang$^{*1}$, Zhaoheng Ni$^{*1}$, \\
  Guangzhi Sun$^{2}$, Pingchuan Ma$^{1}$, Ruizhe Huang$^{\dagger 3}$, Vineel Pratap$^{1}$, Yuekai Zhang$^{4}$, \\
  Anurag Kumar$^{1}$, Chin-Yun Yu$^{5}$, Chuang Zhu$^{4}$, Chunxi Liu$^{\dagger}$, Jacob Kahn$^{1}$, \\ Mirco Ravanelli$^6$, Peng Sun$^{4}$, Shinji Watanabe$^8$, Yangyang Shi$^{1}$, Yumeng Tao$^{\dagger}$, \\
    Robin Scheibler$^{7}$, Samuele Cornell$^{8}$, Sean Kim$^{\dagger}$, Stavros Petridis$^{1}$
  \end{tabular}
   \thanks{$^*$ Equal contribution.}
   \thanks{$^\dagger$ Work done while at Meta.}
}

\address{
   $^1$Meta,
   $^2$University of Cambridge,
   $^3$Johns Hopkins University, \\
   $^4$NVIDIA,
   $^5$Queen Mary University of London,
   $^6$Concordia University, \\
    $^7$LY Corporation,
    $^8$Carnegie Mellon University
}

\newcommand{\code}[1]{\texttt{#1}}

\copyrightnotice{979-8-3503-0689-7/23/\$31.00~\copyright2023 IEEE}

\begin{document}
\ninept
\maketitle

\begin{abstract}
TorchAudio is an open-source audio and speech processing library built for PyTorch. It aims to accelerate the research and development of audio and speech technologies by providing well-designed, easy-to-use, and performant PyTorch components. Its contributors routinely engage with users to understand their needs and fulfill them by developing impactful features. Here, we survey TorchAudio's development principles and contents and highlight key features we include in its latest version (2.1): self-supervised learning pre-trained pipelines and training recipes, high-performance CTC decoders, speech recognition models and training recipes, advanced media I/O capabilities, and tools for performing forced alignment, multi-channel speech enhancement, and reference-less speech assessment. For a selection of these features, through empirical studies, we demonstrate their efficacy and show that they achieve competitive or state-of-the-art performance.

\end{abstract}


\section{Introduction}

With the rapid advancement and increasing pervasiveness of machine learning technologies, usage of open-source toolkits such as Tensorflow~\cite{tensorflow2015-whitepaper} and PyTorch~\cite{NEURIPS2019_9015} for developing machine learning applications has grown significantly. Many modern machine learning applications interface with modalities such as vision, text, and audio. Building such applications, however, requires modality-specific functionality not covered by said general-purpose toolkits.

To address the need for audio and speech facilities in particular, the TorchAudio library has been developed~\cite{9747236}. TorchAudio supplements PyTorch with easy-to-use and performant components for developing audio and speech machine learning models. As a natural extension of PyTorch to the audio domain, TorchAudio embodies many of the same design principles that PyTorch does. Its components support automatic differentiation to facilitate building neural networks and training them end to end. It supports GPU acceleration, which can greatly improve training and inference throughput. It emphasizes composability, simple interfaces shared with PyTorch, and minimal dependencies to allow for easily integrating its components into any application that uses PyTorch.

TorchAudio has been widely adopted and actively developed by the PyTorch community, with its Github development statistics having grown substantially since Version 0.10 was presented in~\cite{9747236} (Table~\ref{tab:usage}). The dramatic increase in the number of repositories that depend on TorchAudio in particular strongly reaffirms TorchAudio's usefulness to the community and success.


This paper begins by summarizing TorchAudio's design principles and contents. It then expounds significant new features that have been introduced since Version 0.10~\cite{9747236}, covering self-supervised learning (Wav2Vec 2.0~\cite{baevski2020wav2vec}, HuBERT~\cite{hsu2021hubert}, XLS-R~\cite{conneau2020unsupervised}, WavLM~\cite{chen2022wavlm}), automatic speech recognition (CTC decoder~\cite{10.1145/1143844.1143891}, Conformer~\cite{gulati20_interspeech}, Emformer~\cite{9414560}, audio-visual speech recognition [AV-ASR]), advanced media I/O, CTC-based forced alignment, multi-channel speech enhancement components, and reference-less speech assessment, of which several are technically novel, e.g. real-time AV-ASR, Emformer, CUDA-based CTC decoder, and CUDA-based forced alignment API. It concludes by presenting experimental results for the new features, which demonstrate that they are effective and achieve or exceed parity in run-time efficiency or output quality with public implementations.

\begin{table}
    \centering
    \setlength{\tabcolsep}{3pt}
    \begin{tabular}{l|ccccc}
\toprule
Version & Contrib. & Commits & Stars & Forks & Dep. repos \\
\midrule
0.10 (Sep 2021) & 144 & 1,013 & 1,428 & 351 & 5,420 \\
2.1 (Jul 2023) & 204 & 2,154 & 2,149 & 585 & 31,173 \\
\bottomrule
\end{tabular}
    \caption{TorchAudio's Github activity statistics, covering unique contributors, commits, stars, forks, and repositories depending on TorchAudio. Statistics for Version 0.10 gleaned from \url{web.archive.org} and~\cite{9747236}.}
    \label{tab:usage}
\end{table}

\section{Related work}
\label{sec:rel}

Several popular open-source toolkits implement lower-level audio operations such as I/O, spectrogram generation, and data augmentations. Just as librosa~\cite{mcfee2015librosa} is one such library for Numpy~\cite{harris2020array} and tfio.audio for Tensorflow, TorchAudio is one such library for PyTorch. The broad applicability of TorchAudio’s data componentry has made it effective in serving more specialized audio data representation libraries such as Lhotse~\cite{elasko2021LhotseAS}, which provides abstractions and utilities that streamline data preparation for downstream audio tasks.

Many higher-level audio and speech machine learning toolkits exist in the PyTorch ecosystem, e.g. ESPnet~\cite{watanabe2018espnet}, SpeechBrain~\cite{ravanelli2021speechbrain}, fairseq~\cite{ott2019fairseq}, and NeMo~\cite{kuchaiev2019nemo}. These toolkits provide ready-to-use models, training recipes, and components covering audio and speech tasks such as text to speech, speech recognition, speech translation, and speech enhancement. As the aforementioned audio operations are fundamental to such tasks, all of these toolkits rely on TorchAudio.

In addition to lower-level audio components, TorchAudio also provides some of the features offered by these higher-level toolkits. For instance, TorchAudio includes task-specific components such as decoders for speech recognition, multichannel functions, and model architectures, as well as ready-to-use models and training recipes. That being said, as far as such features are concerned, TorchAudio is distinguished from many of these other toolkits in its focus on \emph{stable and established technologies} over the cutting edge. For example, rather than maintaining an extensive model repository and continually updating it with the latest state-of-the-art models, we aim to curate a smaller selection of key models and training recipes to demonstrate the use of TorchAudio's components and serve as reliable references. Ultimately, we intend for TorchAudio to be first and foremost a library of established components, which allows it to {\it complement rather than compete with other toolkits in the PyTorch ecosystem}.

\section{Library principles}
TorchAudio firmly adheres to several design principles, which we distill from~\cite{9747236} and clarify.

\paragraph*{Extend PyTorch to audio.}
TorchAudio aims to be PyTorch for the audio domain. Its components compose PyTorch operations, share the same abstractions and Tensor-based interfaces with PyTorch, and support foundational PyTorch features such as GPU acceleration and automatic differentiation. Moreover, its only required dependency is PyTorch. As a consequence, it behaves as a natural extension of PyTorch, and its components integrate seamlessly with PyTorch applications.


\paragraph*{Be easy to use.}
TorchAudio is intuitively designed. Each component is implemented closely following C++, Python, and PyTorch best practices.

It is easy to install. TorchAudio’s binaries are distributed through standard Python package managers PyPI and Conda and support major platforms Linux, macOS, and Windows. Optional dependencies are similarly installable via standard package managers. For users who want to use their own custom logic, building TorchAudio from source is straightforward\footnote{\fontsize{6.5}{5}\selectfont \url{https://github.com/pytorch/audio/blob/main/CONTRIBUTING.md}}.

It is extensively documented. TorchAudio’s official website\footnote{\scriptsize \url{https://pytorch.org/audio/}} comprehensively covers installation directions and the library’s public APIs. Moreover, a wide array of tutorials covering basic and advanced library usages are available on the website and Google Colab. Such resources educate users of all levels of familiarity with audio and speech technologies on how to best use TorchAudio to address their needs.

\paragraph*{Favor stability.}
TorchAudio tends towards mature techniques that are broadly useful. It offers implementations of models and operations that are or will soon become standards in the field. New features are released following a prototype-beta-stable progression to allow users to preview them without disrupting the official releases. Backwards compatibility breaking changes are released after a minimum of two releases to give users ample time to adapt their use cases. 12,000+ test cases and continuous integration workflows run through Github Actions ensure that the APIs work as expected.

\paragraph*{Promote accessibility.} TorchAudio is an open source library. Its entire source code is available on Github\footnote{\scriptsize \url{https://github.com/pytorch/audio}}, where contributions and feedback are encouraged from all. To enable usage in as many contexts as possible, TorchAudio is released under the permissive BSD-2 license.

\section{New Features}
\label{sec:new-feature}
Relative to Version 0.10~\cite{9747236}, TorchAudio 2.1 includes many significant new features. We elaborate on several of these below. Note that some of these features are technically novel and the first of their kind, e.g. the first AV-ASR model to be capable of real-time inference on CPU, the first public implementation of streaming-capable transformer-based acoustic model Emformer, the first CUDA-based CTC beam search decoder, and the first CUDA-based forced alignment API.

\paragraph*{Self-supervised learning.}
Self-supervised learning (SSL) approaches have consistently improved performance for downstream speech processing tasks. While S3PRL~\cite{yang2021superb} focuses on supporting downstream tasks and benchmarking, TorchAudio focuses on upstream models by providing reliable and production-ready pre-trained models and training recipes. TorchAudio now provides models and pre-trained pipelines for Wav2Vec 2.0~\cite{baevski2020wav2vec}, HuBERT~\cite{hsu2021hubert}, XLS-R~\cite{conneau2020unsupervised}, and WavLM~\cite{chen2022wavlm}. Each pre-trained pipeline relies on the weights that the corresponding original model uses and thus produces identical outputs. Moreover, each pipeline is easy to use, simply expecting users to call a single method to retrieve a pre-trained model. To facilitate production usage, TorchAudio's model implementations support TorchScript and PyTorch-native quantization and leverage PyTorch 2.0's Accelerated Transformers\footnote{\fontsize{6.5}{5}\selectfont{\url{https://pytorch.org/blog/accelerating-large-language-models/}}} to speed up training and inference.


TorchAudio also provides end-to-end training recipes that allow for pre-training and fine-tuning HuBERT models from scratch. The training recipes have minimal dependencies beyond PyTorch and TorchAudio and are modularly implemented entirely in imperative code, which makes them conducive to customization and integration with other training flows, as their adoption by other frameworks such as ESPnet~\cite{chen2023reducing} demonstrates.


\paragraph*{CTC decoder.}
Beam search is an efficient algorithm that has been used extensively for decoding speech recognition (ASR) model outputs and remains a fast and lightweight alternative to model-based decoding approaches. We have added a CTC beam search decoder that wraps Flashlight Text’s~\cite{kahn2022flashlight} high performance beam search decoder in an intuitive and flexible Python API. The decoder is general purpose, working for both lexicon and lexicon-free decoding as well as various language model types, including KenLM~\cite{10.5555/2132960.2132986} and custom neural networks, and is easily adaptable to different model outputs.

We have also introduced a CUDA-based CTC beam search decoder. By parallelizing computation along the batch, hypothesis, and vocabulary dimensions, it can achieve much higher decoding throughputs than the CPU-based implementation, which we demonstrate in Section~\ref{sec:cuctc}. To our knowledge, the implementation is the first and only publicly available CUDA-compatible CTC decoder.

\paragraph*{Conformer.}
Conformer is a transformer-based acoustic model architecture that has achieved state-of-the-art results for ASR~\cite{gulati20_interspeech, guo2021recent}. We have developed a PyTorch-based implementation of Conformer and published an RNN-Transducer ASR training recipe that uses it. Using the recipe, we produced a model that achieves word-error-rate (WER) parity with comparable open-source implementations, which will be discussed in Section \ref{sec:conformer}

\paragraph*{Emformer.}
Emformer is a streaming-capable efficient memory transformer-based acoustic model~\cite{9414560}. For on-device streaming ASR applications, it has demonstrated state-of-the-art performance balancing word error rate, latency, and model size. Moreover, because it applies a novel parallel block processing scheme for training, it can be trained very efficiently. We have introduced an implementation of Emformer matching that described in~\cite{9414560} along with an Emformer transducer ASR training recipe and pre-trained inference pipeline. Our implementation is the first to be publicly available, and it has been adopted and extended by icefall\footnote{\url{https://github.com/k2-fsa/icefall/tree/master/egs/librispeech/ASR}}.

\paragraph*{Streaming AV-ASR.} AV-ASR involves transcribing text from audio and video. The vast majority of work to date~\cite{serdyuk2022transformer, DBLP:journals/corr/abs-2201-02184, ma2023auto} has focused on developing non-streaming AV-ASR models; studies on streaming AV-ASR, i.e. transcribing text from audio and video streams in real time, are comparatively limited~\cite{DBLP:journals/corr/abs-2211-02133}. Auto-AVSR~\cite{ma2023auto} is an effective approach to scale up audio-visual data, which enables training more accurate and robust speech recognition systems. We extend Auto-AVSR to real-time AV-ASR and provide an example Emformer transducer training pipeline that incorporates audio-visual input. As far as we know, the AV-ASR model is the first to be capable of real-time inference on CPU.


\paragraph*{Advanced media I/O.}
We have added advanced media processing capabilities to TorchAudio. Class \code{StreamReader} can decode not only audio but also images and videos to PyTorch tensors. Similarly, \code{StreamWriter} can encode tensors as audio, images, and videos. Both support streaming processing as well as applying transforms such as resampling and resizing. They are capable of interfacing with numerous sources and destinations, including file paths and objects, network locations, and devices, e.g. microphones and webcams. Using these features, one can for instance stream audio chunk by chunk from a remote video file and process the corresponding tensors in an online fashion.

We convey the simplicity and versatility of the API via code samples. Figure~\ref{fig:stream_reader_body} instantiates \code{StreamReader} specifying the data source to be a network location, configures output audio and video streams, and iterates over tensors representing chunks of audio and video streamed from the output. Appendix~\ref{appendix:examples} provides additional examples that illustrate how to read from media devices and write to a Real-Time Messaging Protocol server.

\begin{figure}
\centering
\input{figures/stream_reader_figure_body}
\caption{\code{StreamReader} usage example.}
\label{fig:stream_reader_body}
\end{figure}





Furthermore, \code{StreamReader} and \code{StreamWriter} can leverage hardware video processors available on NVIDIA GPUs to greatly accelerate decoding and encoding.

\paragraph*{CTC-based forced alignment.}

We have added support for forced alignment generation, which computes frame-level alignments between audio and transcripts using a CTC-trained neural network model~\cite{kurzinger2020ctc}. The function \code{forced\_align} is compatible with both CPU~\cite{kahn2022flashlight,pratap2019wav2letter++} and GPU~\cite{pratap2023scaling}, providing flexibility to users. The GPU implementation is highly scalable and enables efficient processing of long audio files, and represents the first publicly available GPU-based solution for computing forced alignments.

We provide a tutorial that demonstrates how to effectively use the API. The tutorial also explains how to perform forced alignment for more than 1000 languages using the CTC-based alignment model from the Massively Multilingual Speech (MMS) project~\cite{pratap2023scaling}.

\paragraph*{Multi-channel speech enhancement.} Multi-channel speech enhancement aims to remove noise and interfering speech from multi-channel audio by leveraging spatial properties. Relative to single-channel speech enhancement, multi-channel speech enhancement can produce higher-quality outputs and further enhance the performance of downstream tasks such as ASR~\cite{haeb2019speech}.

Estimating time-frequency masks and applying them to Minimum Variance Distortionless Response (MVDR) beamforming is an established technique capable of robustly improving multi-channel speech enhancement~\cite{araki2016spatial,erdogan2016improved,zhang2021adl}. To support such work, we have implemented a time-frequency mask prediction network and an MVDR beamforming module along with a corresponding training recipe in TorchAudio.


\paragraph*{Reference-less speech assessment.} Speech assessment is essential for developing speech enhancement systems. Existing metrics require either human listening tests, e.g. Mean Opinion Score (MOS), which are expensive and unscalable, or reference clean speech, e.g. Short-Time Objective Intelligibility (STOI), Perceptual Evaluation of Speech Quality (PESQ), scale-invariant signal-to-distortion ratio (Si-SDR), which are impractical for real-world usage.

To address the limitations of such metrics, we have introduced TorchAudio-Squim~\cite{kumar2023torchaudio} — TorchAudio-Speech QUality and Intelligibility Measures — which comprises two neural network based models: one for predicting objective metrics (STOI, wideband PESQ, Si-SDR), and one for predicting subjective metrics (MOS), without reference clean speech.
Broadly speaking, this speech assessment feature establishes a protocol for evaluating speech enhancement without needing any reference signals. We present a case study of its effectiveness in Section~\ref{sec:squim-exp}.

\section{Empirical evaluations}
\label{sec:exp}

We demonstrate the utility of TorchAudio’s new features via studies.

\subsection{Self-supervised learning}
For the HuBERT recipes, we follow the methodology described in~\cite{hsu2021hubert} of first pre-training a model and then fine-tuning it. To pre-train the model, we run two iterations of training. The first iteration trains a HuBERT model on the 960-hour LibriSpeech dataset for 250K steps, with the output targets being clusters mapped from masked frames by a 100-cluster k-means model trained on MFCC features extracted from the dataset. The second iteration trains another HuBERT model on the dataset for 400K steps, with the output targets being clusters assigned by a 500-cluster k-means model trained on intermediate feature representations generated by the first iteration’s model. Then, we fine-tune this final pre-trained model on the 10-hour LibriLight dataset with CTC loss.


Table~\ref{hubert_asr} shows WERs produced in~\cite{hsu2021hubert} and~\cite{chen2023reducing} by evaluating the fine-tuned "Base" model described in the original publication~\cite{hsu2021hubert} on LibriSpeech's test subsets, alongside WERs produced using the same model trained via our recipe and the same decoding strategies. The results validate that our HuBERT training recipes are capable of producing models of quality similar to those described in~\cite{hsu2021hubert} and~\cite{chen2023reducing}. These along with the aforementioned modularity and usability benefits make the models and training recipes particularly promising for users to build upon. Indeed, Chen et al.~\cite{chen2023reducing} adopt TorchAudio’s HuBERT implementation and fine-tuning recipe applying slightly different training approaches, e.g. different k-means training strategies and mixed-precision training with brain floating-point (bfloat16), and achieve better performance than the original (7.6\% and 7.4\% relative WER improvement on test-clean and test-other subsets) while consuming far fewer GPU hours.

\begin{table}
\centering
\begin{tabular}{l|cccc}
\toprule
Subset & Greedy & Greedy in~\cite{chen2023reducing} & 4-gram & 4-gram in~\cite{hsu2021hubert} \\
\midrule
test-clean & 10.9  & 10.5 & 4.4 & 4.3 \\
test-other & 17.8 & 17.6 & 9.5 & 9.4\\
\bottomrule
\end{tabular}
\caption{WER (\%) results of HuBERT fine-tuning model on test subsets of the LibriSpeech dataset. Greedy refers to greedy search decoding and 4-gram to beam search decoding with a 4-gram language model.}
\label{hubert_asr}
\end{table}

\subsection{CTC decoder}
\begin{figure}
    \includegraphics[width=\linewidth]{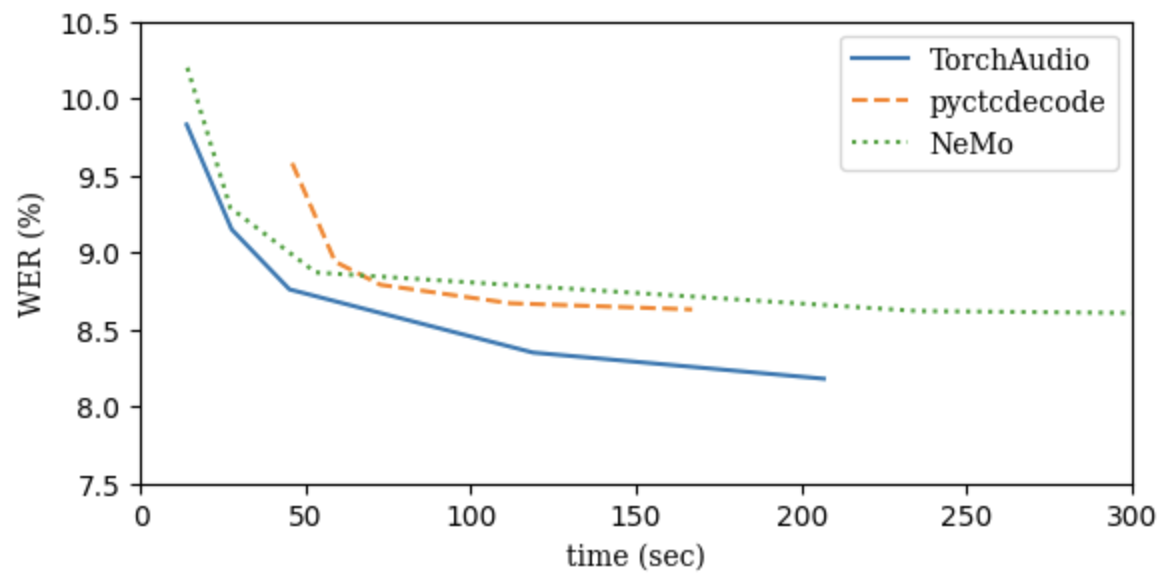}
    \caption{Comparison of WER (\%) vs. time (s) for CPU beam search decoding for TorchAudio, pyctcdecode, and NeMo.}
    \label{fig:cpuctc}
\end{figure}

\paragraph*{CPU CTC decoder.}
The experiments in Figure~\ref{fig:cpuctc} are conducted on LibriSpeech's test-other set on a Wav2Vec 2.0 base model trained on 100 hours of audio. Decoding uses the official KenLM 4-gram LM and takes place on a single thread Intel\textsuperscript{\textregistered} Xeon\textsuperscript{\textregistered} E5-2696 v3 CPU. Because different decoder libraries support different parameters and have different underlying implementations, we first do a sweep for each decoder library for its baseline hyperparameters, and then run decoding with increasing beam sizes for additional data points. We display the wall-clock-time-WER relationship with pyctcdecode\footnote{\scriptsize \url{https://github.com/kensho-technologies/pyctcdecode}} and NeMo~\cite{kuchaiev2019nemo}, where the time in seconds is for decoding the entirety of the test-other dataset. The results show that, for a given target WER, TorchAudio's decoder runs in less time than the baselines. TorchAudio also supports a wider variety of customizable parameters for better hyperparameter tuning and overall WERs.

\paragraph*{CUDA CTC decoder.}
\label{sec:cuctc}
The experiments in Table~\ref{tab:cuctc} are conducted on LibriSpeech's test-other set using a single V100 GPU and Intel\textsuperscript{\textregistered} Xeon\textsuperscript{\textregistered} E5-2698 v4 CPU. For both recipes, a batch size of 4 and a beam size of 10 were applied. The CUDA CTC Decoder uses a CUDA kernel to implement the blank collapse method in~\cite{jung2022blank}. By setting the blank frame skip threshold to $0.95$, the decoding speed can be increased by $2.4$ times without sacrificing accuracy. Since the CPU decoder does not support blank collapsing, the CPU decoder's effective blank frame skip threshold is 1.0. For comparability's sake, then, we include results for the CUDA decoder configured with a blank frame skip threshold of 1.0. By way of CUDA's parallelism, the CUDA decoder allows for performing beam search on all tokens at every step. Thus, the CUDA decoder's effective max tokens per step is the vocabulary size, which is $500$ in this experiment. Accordingly, we include results for the CPU decoder configured with a max tokens per step of $500$ to mimic the CUDA decoder's behavior. Our experimental results show that, compared to the CPU decoder, the CUDA decoder achieves a lower WER and N-best oracle WER while increasing decoding speed by a factor of roughly 10.

\begin{table}
    \centering
    \setlength{\tabcolsep}{3pt}
    \begin{tabular}{l|ccccc}
\toprule
Decoder & Configuration & WER (\%) & Oracle WER (\%) & Time (s) \\
\midrule
CUDA & a = 0.95 & 5.81 & 4.11 & 2.57 \\
CUDA & a = 1.0 & 5.81 & 4.09 & 6.24 \\
CPU & b = 10 & 5.86 & 4.30 & 28.61 \\
CPU & b = 500 & 5.86 & 4.30 & 791.80 \\
\bottomrule
\end{tabular}
    \caption{Comparison of decoding performance for TorchAudio's CUDA and CPU CTC decoders. $a$ is the blank frame skip threshold and $b$ the max tokens per step.}
    \label{tab:cuctc}
\end{table}

\subsection{Conformer}
\label{sec:conformer}
\paragraph*{Model architecture.}
Rather than pursuing state-of-the-art performance, our primary goal is to validate TorchAudio's implementations of Conformer, RNN-T loss, and data operations. Accordingly, we adopt an architecture similar to that used in the baseline Conformer transducer recipes in the ESPnet\footnote{\scriptsize \url{https://github.com/espnet/espnet/tree/master/egs2/librispeech/asr1\#conformer-rnn-transducer}} and icefall\footnote{\scriptsize \url{https://github.com/k2-fsa/icefall/blob/master/egs/librispeech/ASR/RESULTS.md\#2022-04-19}} toolkits.
To ensure comparability, we configure the model architecture to be as similar as possible to those of the baselines. As in the baselines, the encoder has a 512-d output, a subsampling rate of 4, a kernel size of 31, and 8 attention heads with a 2048-d feed-forward unit. In total, our model has 87.4M parameters, with the encoder owning 92\% of them. In contrast, ESPnet's has 94.3M parameters, while icefall's has 84.0M. The differences in parameter count stem mostly from small differences in model architecture, which empirically do not significantly impact performance. For instance, whereas the baselines' encoders use positional embeddings, ours does not.

\paragraph*{Training strategy.}
Building upon TorchAudio's base LibriSpeech Conformer transducer recipe, we create two training recipes, with one reproducing ESPnet's recipe and the other icefall's. Both apply online speed perturbation with factors uniformly sampled from $\{0.9, 1.0, 1.1\}$. The former applies SpecAugment~\cite{Park_2019} with parameters $(T, m_T, F, m_F) = (40, 2, 30, 2)$ and omits additive noise. The latter applies SpecAugment with parameters $(100, 10, 27, 2)$ and includes additive noise, which entails sampling a waveform from MUSAN's “noise” and “music” subsets \cite{musan2015} and adding it to a training sample with probability $0.5$ and signal-to-noise ratio (SNR) uniformly sampled from $(15, 30)$ dB.


Both recipes use the Adam optimizer with weight decay factor 2e-6. Our learning rate scheduler is similar to Noam~\cite{Vaswani2017AttentionIA} in warmup (up to 40 epochs) and annealing (starting from epoch 120 with factor 0.96) steps, with the addition of an 80-epoch plateau at value 8e-4.

\paragraph*{Results.}
With comparable model architectures and training setups, our Conformer transducer recipe performs similarly to or better than ESPnet's and icefall's.

Table~\ref{tab:conformer} shows that the performance of our recipe lies between those of the two ESPnet baselines. Compared with the baseline without CTC auxiliary loss, our recipe produces a model that achieves a 4.3\%/7.8\% relative improvement on test-other/clean. We note, however, that including the auxiliary loss allows the ESPnet recipe to achieve a 10.8\%/9.7\% relative improvement on test-other/test-clean. With the same SpecAugment policy and usage of additive noise, our model performs similarly to icefall's on test-clean and outperforms it by 9.1\% on test-other (Table~\ref{tab:conformer_2}).

\begin{table}
    \centering
    \setlength{\tabcolsep}{3pt}
    \begin{tabular}{l|ccccc}
\toprule
 & & & \multicolumn{2}{c}{WER (\%)} \\
Toolkit & \# param. & Vocab. size & test-clean & test-other \\
\midrule
ESPnet & 91.8M/94.3M* & 5000 & 3.1/2.8* & 7.4/6.6* \\
TorchAudio & 87.4M & 1024 & 2.89 & 7.08 \\
\bottomrule
\end{tabular}


    \caption{Comparison of Conformer transducer recipe model performance between ESPnet and TorchAudio. *With CTC auxiliary loss.}
    \label{tab:conformer}
\end{table}

\begin{table}
    \centering
    \setlength{\tabcolsep}{3pt}
    \begin{tabular}{l|ccccc}
\toprule
& & & \multicolumn{2}{c}{WER (\%)} \\
Toolkit & \# param. & Vocab. size & test-clean & test-other \\
\midrule
icefall & 84.0M & 500 & 2.59 & 6.15 \\
TorchAudio & 87.4M & 1024 & 2.54 & 5.59 \\
\bottomrule
\end{tabular}

    \caption{Comparison of Conformer transducer recipe model performance between icefall and TorchAudio.}
    \label{tab:conformer_2}
\end{table}

\subsection{Streaming AV-ASR}

\paragraph*{Datasets\footnote{\scriptsize All data collection and processing performed at Imperial College London.}.}
In this study, we use the LRS3 dataset~\cite{afouras2018lrs3}, which consists of~151,819 TED Talk video clips totaling 438 hours. Following~\cite{ma2023auto}, we also include English-speaking videos from AVSpeech (1,323 hours)~\cite{DBLP:journals/tog/EphratMLDWHFR18} and VoxCeleb2 (1,307 hours)~\cite{chung2018voxceleb2} as additional training data along with automatically-generated transcriptions. Our model is fed raw audio waveforms and face region of interests (ROIs). We do not use mouth ROIs as in~\cite{ma2023auto, haliassos2022jointly, DBLP:conf/interspeech/ShiHM22} or facial landmarks or attributes during both training and testing.

\paragraph*{Model architecture and training.}
We consider two configurations: Small with 12 Emformer blocks and Large with 28, with 34.9M and 383.3M parameters, respectively. Each AV-ASR model composes frontend encoders, a fusion module, an Emformer encoder, and a transducer model. We use convolutional frontends~\cite{ma2023auto} to extract features from raw audio waveforms and facial images.
The features are concatenated to form 1024-d features, which are then passed through a two-layer multi-layer perceptron and an Emformer transducer model~\cite{9414560}.
The entire network is trained using RNN-T loss.

\paragraph*{Results.}

Non-streaming evaluation results on LRS3 are presented in Table~\ref{non_streaming_evaluation_on_lrs3}. Our audio-visual model with an algorithmic latency~\cite{9414560} of 800 ms ($160\text{ ms}+1280\text{ ms} \times 0.5$) yields a WER of 1.3\%, which is on par with those achieved by state-of-the-art offline models such as AV-HuBERT, RAVEn, and Auto-AVSR. We also perform streaming evaluation adding babble acoustic noise to the raw audio waveforms at various signal-to-noise ratios. With increasing noise level, the performance advantage of our audio-visual model over our audio-only model grows (Table~\ref{noisy_experiments}), indicating that incorporating visual data improves noise robustness. Furthermore, we measure real-time factors (RTFs) using a laptop with an Intel\textsuperscript{\textregistered} Core\textsuperscript{\texttrademark}  i7-12700 CPU running at 2.70 GHz and an NVIDIA 3070 GeForce RTX 3070 Ti GPU. To the best of our knowledge, this is the first AV-ASR model that reports RTFs on the LRS3 benchmark. The Small model achieves a WER of 2.6\% and an RTF of 0.87 on CPU (Table~\ref{streaming_evaluation_on_lrs3}), demonstrating its potential for real-time on-device inference applications.

\begin{table}[ht]
\centering
\begin{tabular}{c|c|c}
\toprule
Method & Total Hours & WER (\%) \\
\midrule
ViT3D-CM~\cite{serdyuk2022transformer} & 90,000 & 1.6 \\
AV-HuBERT~\cite{DBLP:conf/interspeech/ShiHM22} & 1,759 & 1.4 \\
RAVEn~\cite{haliassos2022jointly} & 1,759 & 1.4 \\
Auto-AVSR~\cite{ma2023auto} & 3,448 & 0.9 \\
Ours & 3,068 & 1.3 \\
\bottomrule
\end{tabular}
\caption{Non-streaming evaluation results for audio-visual models on the LRS3 dataset.}
\label{non_streaming_evaluation_on_lrs3}
\end{table}

\begin{table}[ht]
\centering
\begin{tabular}{c|c|c|c|c|c|c}
\toprule
Type &$\infty$\,dB & 10\,dB & 5\,dB & 0\,dB & -5\,dB & -10\,dB \\
\midrule
A   &1.6 &1.8 & 3.2 &10.9 &27.9 &55.5 \\
A+V &1.6 &1.7 & 2.1 &6.2  &11.7 &27.6 \\
\bottomrule
\end{tabular}
\caption{Streaming evaluation WER (\%) results at various signal-to-noise ratios for our audio-only (A) and audio-visual (A+V) models on the LRS3 dataset under 0.80-second latency constraints.}
\label{noisy_experiments}
\end{table}

\begin{table}[ht]
\centering
\begin{tabular}{c|c|c|c}
\toprule
Model & Device  & WER (\%) & RTF \\
\midrule
Large & GPU  & 1.6 & 0.35 \\
\midrule
 \multirow{2}{*}{Small} & GPU & \multirow{2}{*}{2.6} & 0.33 \\
 & CPU & & 0.87 \\
\bottomrule
\end{tabular}
\caption{Impact of AV-ASR model size and device on WER and RTF. Note that the RTF calculation includes the pre-processing step wherein the Ultra-Lightweight Face Detection Slim 320 model is used to generate face bounding boxes.}
\label{streaming_evaluation_on_lrs3}
\end{table}

\subsection{Multi-channel speech enhancement}
\label{sec:mvdr}
\paragraph*{Datasets.} To validate the efficacy of TorchAudio's MVDR beamforming module, we use the L3DAS22 3D speech enhancement task (Task1) dataset~\cite{guizzo2022l3das22} which contains 80 and 6 hours of audio for training and development, respectively. Each sample in the dataset comprises a far-field mixture recorded by two four-channel ambisonic microphone arrays and the corresponding target dry clean speech and transcript.


~\paragraph*{Model architecture and training.}
Experiments are conducted following the mask-based MVDR beamforming methodology described in~\cite{araki2016spatial}. First, a Conv-TasNet-based mask network is applied to compute the complex-valued spectrum and estimate the time-frequency masks for speech and noise. The mask network consists of a short-time Fourier transform (STFT) layer and a Conv-TasNet model with its feature encoder and decoder removed. Then, the MVDR module is applied to the masks and multi-channel spectrum to produce the beamforming weights. Finally, the beamforming weights are multiplied with the multi-channel STFT to produce a single-channel enhanced STFT from which the enhanced waveform is derived via inverse STFT. We use Ci-SDR~\cite{boeddeker2021convolutive} as the loss function since dry clean signals are generally not aligned with multi-channel inputs in real-world scenarios. Model configurations and training details can be found in~\cite{lu2022towards}.



~\paragraph*{Results.}

We evaluate the impact of the mask-based MVDR beamforming model alongside various baselines on downstream ASR performance. First, we evaluate each model on the test set of the L3DAS22 dataset to generate the corresponding enhanced speech. Then, we evaluate the Conformer transducer model presented in Section~\ref{sec:conformer} on the enhanced speech and compute the WER between the generated transcriptions and the true transcriptions. Separately, we also evaluate a Wav2Vec-2.0-based ASR model on the enhanced speech and dry clean speech and compute the WER between the two sets of generated transcriptions, per the L3DAS22 Challenge's WER metric. The results (Table~\ref{mvdr_results}) imply that the mask-based MVDR model significantly improves ASR performance compared to other methods, validating the efficacy of TorchAudio's MVDR module.

\begin{table}[t]
\centering
\begin{tabular}{l|c|c|c|c}
\toprule
Model & WER (\%) & WER$*$ (\%)\\
\midrule
MIMO-UNet~\cite{guizzo2022l3das22} (baseline) & 9.4 & 25.0\\
FasNet~\cite{luo2019fasnet} & - & 14.2 \\
DCCRN~\cite{hu2020dccrn} & - & 18.8 \\
Demucs v3~\cite{defossez2021hybrid} & - & 15.3 \\
Mask-based MVDR & 3.5 & 5.6\\
\midrule
Noisy Mixture & 11.5 & 46.7\\
Dry Clean & 2.6 & 0.0\\
\bottomrule
\end{tabular}
\caption{WER results for the mask-based MVDR beamforming model and other baselines on the test set of the L3DAS22 dataset.  WER corresponds to the WER computed for the Conformer transducer model and WER$*$ to the L3DAS22 Challenge's WER metric.}
\label{mvdr_results}
\end{table}

\subsection{Reference-less speech assessment}
\label{sec:squim-exp}
As discussed in Section~\ref{sec:mvdr}, it can be challenging to compute signal-level speech enhancement metrics (e.g., Si-SDR) in real-world scenarios since obtaining aligned dry clean signals is difficult. Here, we conduct a case study of TorchAudio-Squim's utility in evaluating enhanced signals assuming such scenarios. Using TorchAudio-Squim, we estimate STOI, PESQ, and Si-SDR for the enhanced speech generated in Section~\ref{sec:mvdr}. Table~\ref{squim_results} suggests that the scores predicted by TorchAudio-Squim are consistent with actual speech quality and intelligibility.

By jointly leveraging TorchAudio's mask-based MVDR beamforming model, Conformer transducer model, and TorchAudio-Squim, we show that one can perform multi-channel speech enhancement, ASR, and speech quality assessment all within TorchAudio.

\begin{table}[t]
\centering
\setlength\tabcolsep{2.1pt}
\begin{tabular}{l|c|c|c|c|c|c}
\toprule
& \multicolumn{3}{c|}{Real metrics} & \multicolumn{3}{c}{TorchAudio-Squim} \\
Model & WER & PESQ & Ci-SDR & STOI & PESQ & Si-SDR\\
\midrule
MIMO-UNet~\cite{guizzo2022l3das22} & 9.4 & 1.93 & 8.26 & 0.90 & 1.83 & 10.33\\
mask-based MVDR & 3.5 & 2.46 & 19.00 & 0.92 & 2.25 & 15.95 \\
\midrule
Noisy Mixture & 11.5 &  1.21 & 1.87 & 0.67 & 1.22 & -1.70 \\
Dry Clean & 2.6 & 4.64 & $\infty$ & 0.99 & 3.56 & 23.93\\
\bottomrule
\end{tabular}
\caption{WER, PESQ, Ci-SDR, and SQUIM metric results for the mask-based MVDR beamforming model and other baselines on the test set of the L3DAS22 dataset.}
\label{squim_results}
\end{table}

\section{Conclusion}
TorchAudio 2.1 offers many compelling audio and speech machine learning components. Not only are its components well-designed and easy to use, but they are also effective and performant, as corroborated by our empirical results. Consequently, the library establishes a sound basis for future work in alignment with its ultimate goal of accelerating the advancement of audio technologies, and we look forward to seeing what its incredible community of users will achieve with it next.

\section{Acknowledgements}
We thank all of TorchAudio’s contributors on Github, including Grigory Sizov, Joel Frank, Kuba Rad, Kyle Finn, and Piotr Bialecki. We thank Andrey Talman, Danil Baibak, Eli Uriegas, Nikita Shulga, and Omkar Salpekar for helping with TorchAudio’s releases. We thank Abdelrahman Mohamed, Buye Xu, Daniel Povey, Didi Zhang, Donny Greenberg, Hung-yi Lee, Laurence Rouesnel, Matt D’Zmura, Mei-Yuh Hwang, Soumith Chintala, Thomas Lunner, Wei-Ning Hsu, and Xin Lei for the many valuable discussions.



{
\footnotesize
\bibliographystyle{IEEEbib}
\bibliography{main}
}

\clearpage

\appendix

\section{Usage examples}
\label{appendix:examples}
\begin{figure}[h]
\centering
\input{figures/stream_reader_figure_appendix}
\caption{Sample code that uses \code{StreamReader} to process streaming input from media devices.}
\label{fig:stream_reader}
\end{figure}

\begin{figure}[h]
\centering
\input{figures/stream_writer_figure}
\caption{Sample code that uses \code{StreamWriter} to live-stream audio-visual data via Real-Time Messaging Protocol server.}
\label{fig:stream_writer}
\end{figure}

\end{document}